# A LINEAR MODELING APPROACH ON LTGP SIGNALS AND SEISMIC ACTIVITY

Dimitrios Dechouniotis[1] and Apostolos Ifantis [2]

[1] School of Electrical and Computer Engineering, National Technical University of Athens, Athens, Greece
[2] Department of Electrical Engineering, Technical Education Institute of Western Greece, Patras, Greece

## ABSTRACT

*This study illustrates presents a set of the Long – Term Geoelectric Potential (LTGP) measurements that are collected for experimental investigation in Western Greece during a five–year period (1993 – 1997). During this period, many major destructive earthquake events occurred that caused human casualties and extended material damages. The collection and processing of geoelectric measurements was done by an automated data acquisition system at the Seismological Laboratory of the University of Patras, Greece. This novel study considers seismic activity of this area as a typical linear dynamic system and the dynamic relationship between the magnitude of earthquakes and Long – Term Geoelectric Potential signals is inferred by the Recursive Least Square algorithm. The results are encouraging and show that linear dynamic systems, which are widely used in modern control theory, can describe efficiently the dynamic behavior of seismic activity and become a useful interpretative tool of seismic phenomenon.*

## 1. INTRODUCTION

Earthquakes' disastrous effects on human lives has motivated scientists putting more effort on earthquake prediction based on precursor events than on earthquake forecasting in a probabilistic manner. A great variety of precursor signals have been proposed through years, among which geoelectromagnetic ones, captured through geoelectrical measurements over a broad frequency spectrum, have been promising on an eventual prediction of damaging earthquakes [1], [2], [3]. Earthquake event occurrence can be considered as the outcome of preseismic geotectonic variations reflected on the geoelectric potential. In this way, the behavior of Long-Term Geoelectrical Potential (LTGP) signals can be correlated to an upcoming earthquake event of great importance. Research on this direction has given interesting results.

The chaotic behavior of the LTGP as a seismic event precursor signal has been explored and confirmed using Lyapunov exponents and Takens estimator [4], [5]. The fractal analysis of the Ultra-Low Frequency LTGP has revealed that strong earthquakes are shown to be preceded by a decrease of the spectral power law exponent approaching unity [6], [7]. Also, in an experimental analysis of seismic activity in Western Greece, a possible positive correlation of the Hurst exponent of LTGP difference signal to the changes of the spectral power-law exponent of the seismic activity has been found prior to major events [8].

LTGP power spectrum features relation to seismic activity has also been extensively investigated. Initially, the Short Time Fourier Transform has been used [9]. The encouraging results motivated scientists to deploy other advanced time-frequency analysis tools, such as [10] and Teager Huang Transform [11].

In this paper, a linear dynamical system is used to model the dynamic relation between the magnitude of an earthquake and LTGP signals. This type of modeling has been broadly adopted in modern control theory for many different fields, e.g., electrical or mechanical systems, computer and network systems [18], robotics etc. Raw LTGP data were collected at the Earthquake Prediction Section of the University of Patras Seismological Laboratory (UPSL) [12] during a five–year period, 1993–1997, in which significant events occurred close to the observation station. To the best of our knowledge this is the first study that considers the seismic activity as a dynamical system. The produced results are encouraging and show that a linear dynamical system can be an accurate mathematical model of an earthquake.

The rest of this paper is structured as follows; in Section 2 the acquisition system along with the data are described. Section 3 provides the algorithm for identifying the dynamic model of an earthquake. Results are presented in Section 4. Finally, conclusions are drawn.

## 2. DATA ACQUISITION AND MEASUREMENTS

The geological procedures prior to an earthquake event include the application of stress on rocks, which causes the redistribution of pore fluids, fluid flow, and the piezoelectric effect of quartz. These explain the generation of electrical signals precursor to earthquake events. The measurement of the LTGP signals is done using two pairs of Pb-PbCL2 electrodes, perpendicular to each other, in NS (*ch0*) and EW (*ch1*) direction. In each dipole pair, electrodes are placed in a 100m distance. Dipole electrotelluric signals are directed to an electronic VAN device, converted to digital with a sampling rate of 3 samples/min. Channel signals are in further anti-alias filtered using a Butterworth low-pass filter sampled with a 32-bit resolution and transferred to a PC in the control room via a dedicated line for further monitoring and processing. Also, a pen recording system is used for reliability and illustration purposes. The structure of the data acquisition system is presented in Figure 1.

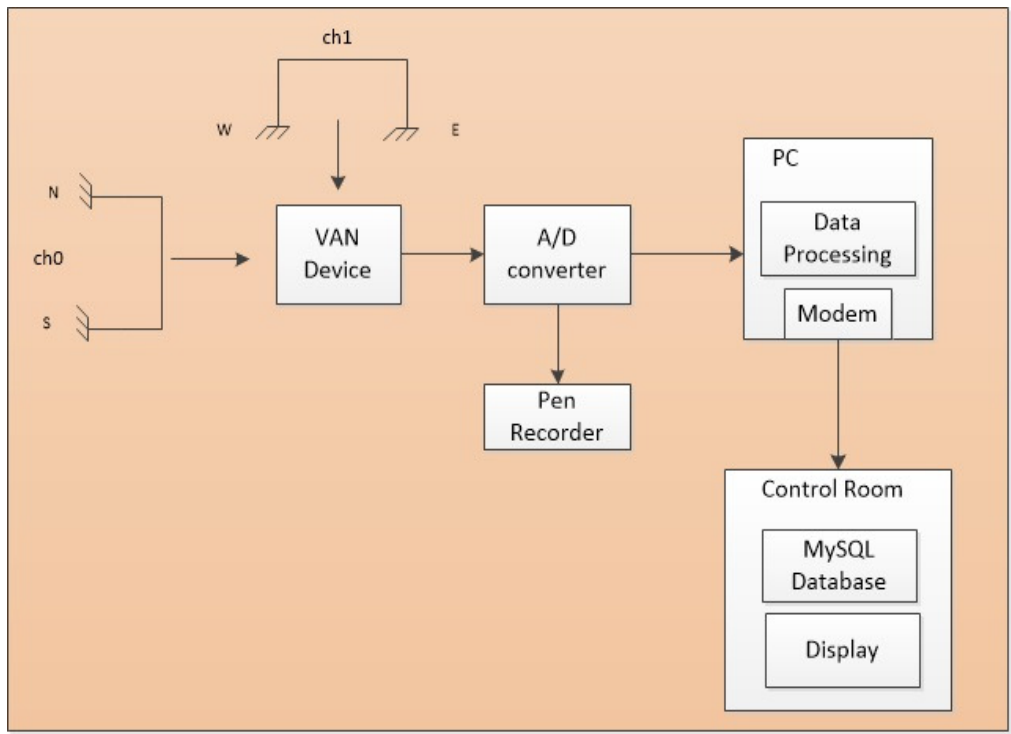

Figure 1. Data Acquisition System

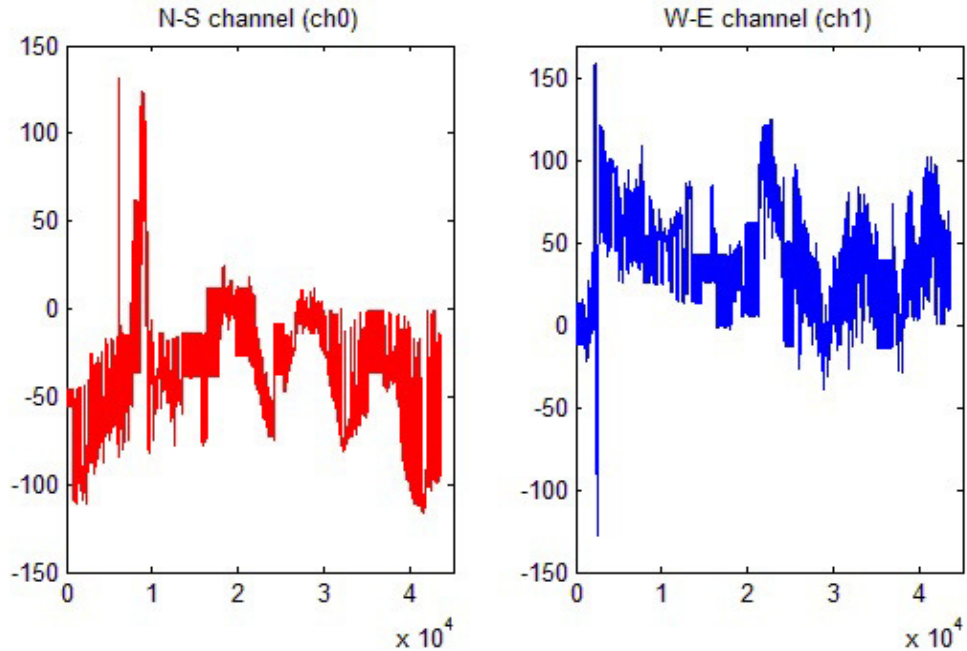

Figure 2. Observed LTGP signals

In order to observe long term variation, the sampling (digitization) rate was in further decreased to 1sample/hr. Thus, for the examination period of 1993-1997 in Western Greece a data set of 43824 points was obtained. The relevant signals of ch0 and ch1 are depicted in Figure 2. Additionally, for the same period, the major earthquake events with magnitude $M_s \geq 4.8$ close to our recording station, as provided by the Institute of Geodynamics of National Observatory of Athens [13], are recording for the exploration of their correlation to the precursor electrotelluric signals. Figure 3 shows the exact location of these events, while Table 1 contains all the necessary information about these major earthquakes.

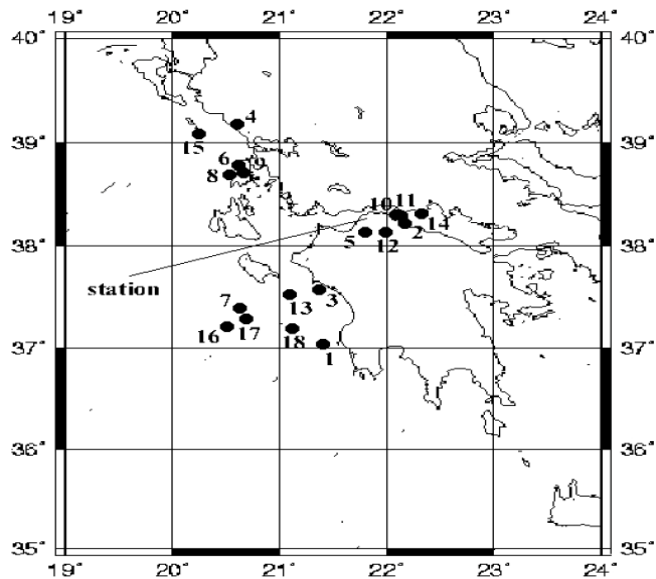

Figure 3. Epicenter of major Earthquakes in Western Greece (1993-1997)

Table 1. Major Earthquakes occurred in Western Greece during the period 1993-1997

| No. | Point | Date [yy/mo/dd] | Distance [km] | Depth [km] | Longitude | Latitude | Magnitude |
| --- | --- | --- | --- | --- | --- | --- | --- |
| 1 | 1543 | 1993/03/05 | 176 | 6.4 | 37.07 | 21.46 | 5.3 |
| 2 | 1863 | 1993/03/18 | 44 | 1.5 | 38.26 | 22.20 | 4.9 |
| 3 | 2055 | 1993/03/26 | 100 | 18.3 | 37.65 | 21.44 | 5.0 |
| 4 | 3960 | 1993/06/13 | 192 | 40.0 | 39.25 | 20.57 | 5.4 |
| 5 | 4700 | 1993/07/14 | 15 | 50.2 | 38.16 | 21.76 | 5.1 |
| 6 | 10082 | 1994/02/25 | 147 | 49.7 | 38.73 | 20.58 | 5.3 |
| 7 | 11304 | 1994/04/16 | 182 | 36.4 | 37.43 | 20.58 | 5.3 |
| 8 | 16744 | 1994/11/29 | 159 | 28.2 | 38.66 | 20.46 | 4.9 |
| 9 | 16783 | 1994/12/01 | 148 | 35.8 | 38.69 | 20.55 | 4.8 |
| 10 | 21505 | 1995/06/15 | 42 | 50.9 | 38.36 | 22.15 | 5.6 |
| 11 | 21506 | 1995/06/15 | 43 | 3.7 | 38.33 | 22.18 | 5.1 |
| 12 | 24030 | 1995/09/28 | 29 | 30.4 | 38.15 | 21.99 | 4.8 |
| 13 | 30064 | 1996/06/06 | 130 | 35.8 | 31.55 | 21.11 | 4.9 |
| 14 | 42453 | 1997/11/05 | 55 | 28.3 | 38.34 | 22.31 | 4.9 |
| 15 | 42616 | 1997/11/12 | 200 | 56.6 | 39.10 | 20.27 | 4.8 |
| 16 | 42757 | 1997/11/18 | 206 | 36.9 | 37.26 | 20.49 | 6.1 |
| 17 | 42758 | 1997/11/18 | 183 | 48.3 | 37.36 | 20.65 | 5.6 |
| 18 | 42759 | 1997/11/18 | 209 | 32.4 | 37.25 | 21.16 | 5.0 |

## 3. LINEAR TIME-INVARIANT STATE SPACE MODEL

In control theory, the dynamic evolution of a process is defined mathematically by a state–space model. This kind of modeling is suitable for both continuous and discrete systems. In both cases, a vector of inputs *u* consists of all control variables and the vector of states *x* contains all variables that describe the dynamic evolution of the system. Generally, a state-space model is linear or non-linear function of the state and input vectors *f(x,u)*. In the case of discrete state-space model, a difference equation denotes how the state variables of the next time interval evolve from the current values of input and state vector, *x(k + 1) = f(x(k),u(k))*. A discrete model can be derived directly from difference equations if they are available. On the contrary, the complexity, linear or not, and the parameters of function *f(x, u)* can be selected so that an identification method produce an explicit model from measurements of input and state variables. System identification is an important area of systems theory [14] that provide black box methods for deriving state–space model form system's measurements. In this study, a Linear Time-Invariant (LTI) model is chosen to describe the earthquake as a dynamic system, which mathematically defined by,

$$x_k = a_{k-n}x_{k-n} + \cdots + a_{k-1}x_{k-1} + b_{k-m}u_{k-m} + \cdots + b_{k-1}u_{k-1} \quad (1)$$

$$x_{k+1} = Ax_k + Bu_k \quad (2)$$

where $x_k \in \mathbb{R}^2$ is the state vector that contains *ch0* and *ch1* signals and $u_k \in \mathbb{R}$ is the input vector that is the magnitude of the earthquake in every kth time interval respectively. The order of the model is the number of the elements of vector $x_k$. In order to increase the accuracy of the model, past values of the input variables and states can be used. However, this increases the complexity of the model. There is always a trade–off between the accuracy and complexity of the model as the order of the model varies. The Recursive Least Square (RLS) algorithm [15] is the most common identification algorithm for identifying linear model of processes in industry. In this paper, this algorithm is chosen to compute the elements of matrices *A* and *B* of the LTI model

International Journal of Chaos, Control, Modelling and Simulation (IJCCMS) Vol.7, No.1, March 2018

due to its simplicity and accuracy. Furthermore, it can be implemented online and does not require significant computational resources.

## 4. VALIDATION AND RESULTS

In order to validate whether an LTI model can effectively capture the dynamics of an earthquake, the acquired data set is separated geographically. In Figure 3, it is obvious that the major earthquakes can be separated in three geographical areas. The first area, named $Area_1$, is around Lefkada island in North-Western Greece (Longitude: 20-21, Latitude: 38-39.5). The second area, called $Area_2$, lies on the North coast of Peloponnese (Longitude: 21.5-22.5 Latitude: 38-39), which is closest to the data acquisition system. The marine area west of Peloponnese (Longitude: 20-22, Latitude: 37-38) is the third area of interest, called $Area_3$.

The effectiveness of RLS algorithm is tested under two different scenarios. Initially, the RLS algorithm is fed with the whole data set to produce a unique LTI model for the whole region of Western Greece and an LTI model for each geographical area. Secondly, we focus on the major earthquakes of Table 1. For each geographical area, a local data set is created by concatenating measurements one week before and after a major earthquake. Then RLS algorithm used these data sets to produce three LTI models for the week before the major earthquake, for the week after the earthquake and for both weeks together for each geographical area. All derived models are evaluated using the Best Fit Rate (BFR) [16], [17], that is a well-known metric of system identification process. In order to produced unbiased results, different data were used for the training and the evaluation process. For both scenarios, we derive second and four-order LTI models. Higher order models do not improve the accuracy of the model. More specifically the second and fourth order LTI models are given by Eq. (3)-(4) respectively,

$$\begin{bmatrix} x1_{k+1} \\ x2_{k+1} \end{bmatrix} = \begin{bmatrix} a_{11} & a_{12} \\ a_{21} & a_{22} \end{bmatrix} \begin{bmatrix} x1_k \\ x2_k \end{bmatrix} + \begin{bmatrix} b_1 \\ b_2 \end{bmatrix} m_k \qquad (3)$$

Where $\begin{bmatrix} x1_k \\ x2_k \end{bmatrix} = \begin{bmatrix} ch0_k \\ ch1_k \end{bmatrix}$,

$$\begin{bmatrix} x1_{k+1} \\ x2_{k+1} \\ x3_{k+1} \\ x4_{k+1} \end{bmatrix} = \begin{bmatrix} a_{11} & a_{12} & a_{13} & a_{14} \\ 1 & 0 & 0 & 0 \\ a_{31} & a_{32} & a_{33} & a_{34} \\ 0 & 0 & 1 & 0 \end{bmatrix} \begin{bmatrix} x1_k \\ x2_k \\ x3_k \\ x4_k \end{bmatrix} + \begin{bmatrix} b_1 \\ 0 \\ b_2 \\ 0 \end{bmatrix} m_k \qquad (4)$$

Where $\begin{bmatrix} x1_k \\ x2_k \\ x3_k \\ x4_k \end{bmatrix} = \begin{bmatrix} ch0_k \\ ch0_{k-1} \\ ch1_k \\ ch1_{k-1} \end{bmatrix}$.

The RLS algorithm uses time-series of inputs and state variables in order to derive the corresponding models. The recorded measurements of ch0 and ch1 with digitization rate one sample per hour are the values of state variables. The input measurements are derived by the earthquakes data set with digitization rate of 1 sample per hour. If more than one earthquake took place during the one-hour interval, the magnitude of the strongest earthquake is used. If there is no earthquake, the measurement of this time interval is set to zero.

International Journal of Chaos, Control, Modelling and Simulation (IJCCMS) Vol.7, No.1, March 2018

As mentioned earlier, at the first scenario a single model corresponds to the whole data set and an LTI model for each area respectively. Eq. (2) is the simplest LTI model that it can be used. A set of 30000 measurements is used for training and the rest data are used for the evaluation of the model.

Figure 4 shows that the RLS algorithm converges quickly to the final values and Figure 5 illustrates that the produced models are very precise. The results for the produced fourth order LTI model are similar and they are omitted for brevity.

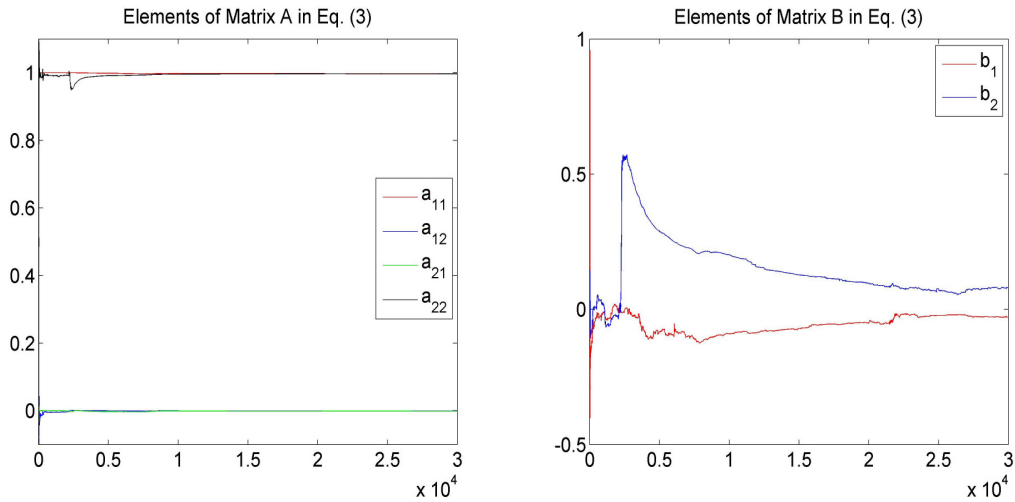

Figure 4. Convergence of Second Order LTI Model

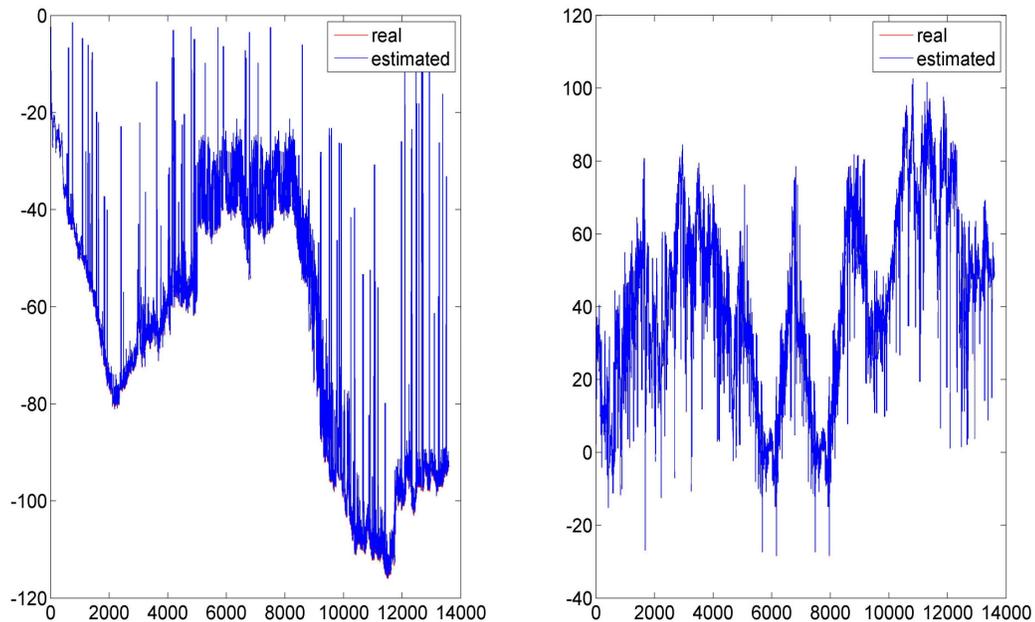

Figure 5. Real Measurements and Produced Data by LTI Model

Table 2. First Scenario: BFR Score of the RLS Algorithm

| Data Set | Second Order Model BFR (%) | Fourth Order Model BFR (%) |
| --- | --- | --- |
| Entire | 85.02 | 85.11 |
| Area1 | 85.66 | 85.79 |
| Area2 | 87.96 | 88.13 |
| Area3 | 86.48 | 86.57 |

Table 2 presents the BFR score of the second and fourth order LTI model for the entire data set and each geographical area respectively. For each area the accuracy of the LTI model is high and independent of the distance from the data acquisition system. As it was expected, the fourth order LTI model is slightly improved than second order model.

At the second scenario, our work investigates the accuracy of LTI model for the major earthquakes. The entire data set for $Area_1$ and $Area_2$ consists of 1685 measurements and the set of $Area_3$ includes 2022 samples. The produced models are derived from a portion of two third of the entire data set for each area and the rest measurements are used for validation. Figures 6-7 present the final value of the elements of matrices $A$ and $B$ of the four order LTI model of $Area_3$. Figure 8 depicts the difference between real measurements and the produced values by the derived LTI model. Although that the training sets are relatively small, the algorithm converges quickly to the final values of the models (Figures 6-7) and its precision is still very high (Figure 8).

Table 3 presents the BFR score of the second and fourth order LTI model for the major earthquakes of each geographical area. The columns of this table correspond to models for one-week period before the earthquake, one week after the earthquake and for both weeks respectively. The upper part of the Table 3 corresponds to second order LTI model and the lower part corresponds to the fourth order model. It is worth-mentioned that models corresponding after the major earthquake or for the two-week period are the most accurate. Also, the fourth order model is slightly precise than the second order model.

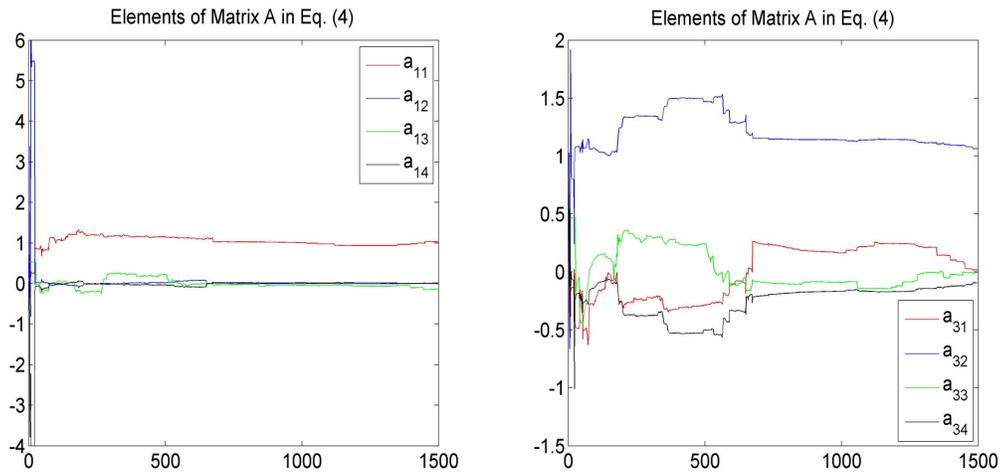

Figure 6. Fourth Order LTI Model's Parameters (Matrix A)

International Journal of Chaos, Control, Modelling and Simulation (IJCCMS) Vol.7, No.1, March 2018

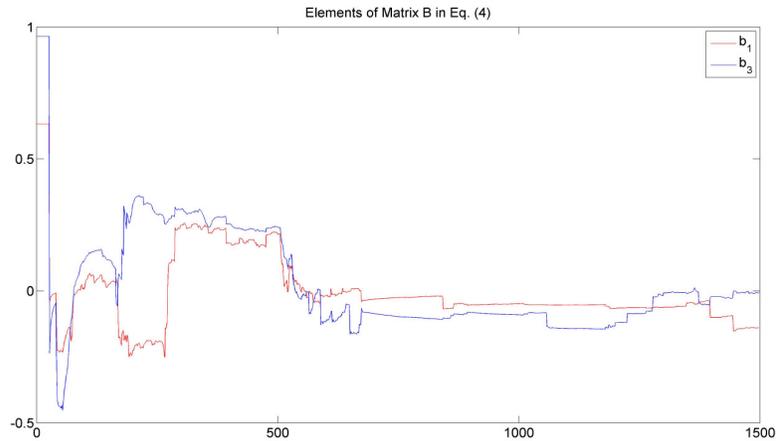

Figure 7. Fourth Order LTI Model's Parameters (Matrix B)

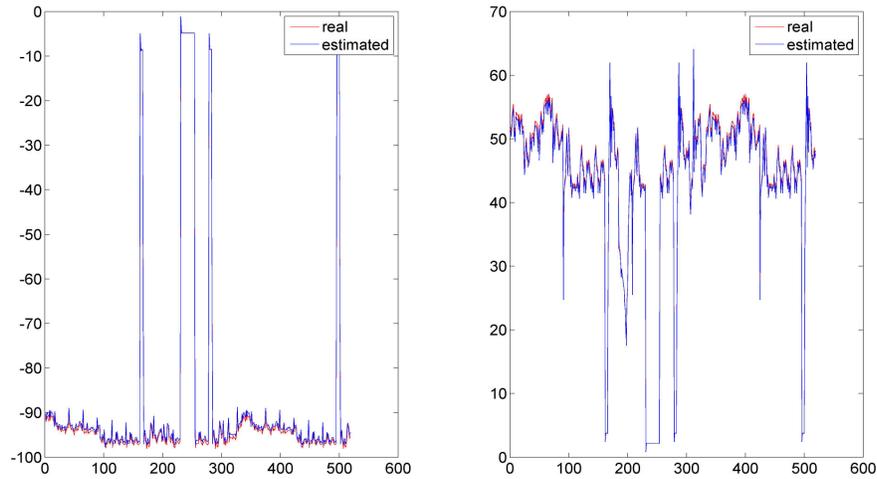

Figure 8. Real Measurements and Produced Data by LTI Model

Table 3 – Second Scenario: BFR Score of the RLS Algorithm

| | Second Order Model BFR (%) | | |
|---|---|---|---|
| | Before Data Set | After Data Set | Entire Data Set |
| $Area_1$ | 82.7 | 84.01 | 84.77 |
| $Area_2$ | 80.02 | 79.79 | 80.99 |
| $Area_3$ | 76.15 | 78.46 | 78 |
| | Fourth Order Model BFR (%) | | |
| $Area_1$ | 83.13 | 83.96 | 84.82 |
| $Area_2$ | 80.45 | 79.99 | 81.09 |
| $Area_3$ | 76.12 | 78.39 | 77.97 |

This study is a "reverse-engineering" method compared with [4]; because the magnitude of the earthquake is the cause and the LTGP signals are the results. From the results of both scenarios, it is inferred that LTI models can satisfactorily describe the dynamic relationship between LTGP signals and the earthquake's magnitude. It is validated that the produced LTI models are accurate independent of the presence of major earthquakes and the distance from the data acquisition system. This study can be used to simulate the seismic activity of a specific area. Furthermore, it can be combined with other prediction tools from modern control theory in order to predict destructive earthquakes using historical data.

## 5. CONCLUSIONS AND FUTURE WORK

To the best of our knowledge this is the first study that investigates the dynamic relation between the magnitude of major earthquakes and LTGP signals. The widely used RLS algorithm is used to derive $2^{nd}$ and $4^{th}$ order discrete-time LTI models and tunes their parameters. The initial results show that the derived models are precise independent of the presence of major earthquakes and the distances from the data acquisition location. This kind of modeling combined with prediction methods from modern control theory seems promising in order to extract useful information about seismic activity of an area and the forecasting of destructive earthquakes.

International Journal of Chaos, Control, Modelling and Simulation (IJCCMS) Vol.7, No.1, March 2018

**AUTHOR**

Dr. Dimitrios Dechouniotis is a Senior Researcher at NETMODE Lab, NTUA Greece. He received his diploma from the Department of Electrical and Computer Engineering (ECE) at the University of Patras, Greece, in 2004, his Master of Science degree from the NTUA, Greece, in 2006, and his PhD degree from the Department of ECE at the University of Patras, in 2014. He was Partner Lecturer at the Electrical Engineering Department of Technical Educational Institute of Western Greece during 2007-2016. His research interests are network monitoring, cloud computing, pervasive computing and control theory.

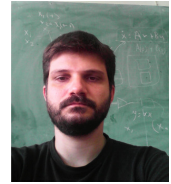

Dr Apostolos Ifantis received the B.S. degree on Physic Sciences in 1976 and the Ph.D. degree in 1994 from the University of Patras, Greece. Currently, he is Professor in the Department of Electrical Engineering at the Technological Educational Institute of West Greece (Patras) and collaborating teaching staff at the University of Patras. Also, he is responsible of the Center of Pre-seismic Signals of the Laboratory of Geophysics and Seismology of the University of Patras. His research interests include automation control systems, data acquisition, signal analysis and digital processing, image processing, data mining and sensors. He has published more than 40 Journal and International Conference papers in the above fields and has over 50 references to his research work.

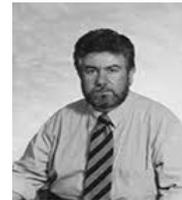